# Neuro-Adaptive Boundary Force Control of Dual One-Link Flexible Arms with Unmodeled Dynamics and Input Constraints

Mahdi Hejrati

*Abstract*— **The primary purpose of this article is to accomplish safe grasping task by means of dual one-link flexible manipulators. In order to design a force-sensor-less force control, the direct force control problem is reduced to common motion control problem, in a way that by satisfying new control objectives the grasping task is established. Afterwards, for the first time in the field of dual one-link flexible manipulators, intelligent control methods are combined with robust control approaches in an effort to; I) accomplish motion control objectives, II) handle uncertainties in the system, and III) consider unknown, mixed input constraints, resulting in NABFC (Neuro-Adaptive Boundary Force Control). Moreover, to deal with unknown model uncertainties as well as unknown input saturation and dead zones, Radial Basic Function Neural-Networks (RBFNNs) are used. In the same way, adaptive control is utilized to estimate unknown parameters. By exploiting Lyapunov's direct method, proper Lyapunov functional and Energy multiplier method are defined to express well-known yet strong stability procedure, which compensates a complex stability procedure proposed in the previous works. In the presence of the designed controller, the presented stability procedure resulted in a uniform ultimate boundedness (UUB) stability for the system. Finally, for comparison aim between the designed controller with other controllers, numerical analysis is used to demonstrate both the excellent performance of the proposed controller and the correctness of the stability analysis outcomes.**

## I. Introduction

In the face of the energy consumption crisis in various industrial applications, utilizing flexible link manipulators can be a feasible alternative. Have been designed to attain good positioning accuracy, prevalent robotic manipulators are fabricated in a massive and bulky way, which leads to low operational speed and needs big actuators. The ensuing consequence of mentioned problems makes these manipulators energy inefficient. On the contrary, because of low stiffness, flexible manipulators are lightweight, which diminishes power utilization.

Despite of all assets, flexible manipulators suffer from lack of accuracy at the end-effector's position due to unwanted vibration of flexible links. Hence, vibration suppression and position control of such manipulators are critical objectives. In the context of one-link flexible manipulators, throughout the last decades, numerous works have been done[1]–[4]. However, having more dexterous manipulators aiming for executing more complex tasks necessitates controlling contact force subjected to object or environment by end-effector. Controlling the grasping force through dual one-link flexible manipulators can prepare more safe conditions to do so. Due to their low inertia, such manipulators can have lower force overshoot that can be used for manipulating vulnerable objects[5].

The mathematical model of these arms consists of hybrid PDE-ODE equations that represents flexible link vibration and motor dynamics along with geometrical constraints. However, although considering PDE model for flexible arms makes analysis of these systems more challenging, it eliminates some problems like spillover instability resulted by reducing the infinite-dimensional model to the finite one[6].

Albeit rigid cooperative manipulators have been exceedingly examined[7]–[9], flexible ones have appealed little attention of researchers. In [10], the grasping task using dual one-link Euler-Bernoulli arms has been studied. To remove unwanted vibration and apply desired grasp-ensuring force to an object a simple boundary controller has been designed. Moreover, Semigroup theory and Hilbert spaces have been employed to reach conditional asymptotic and exponential stabilities. In [11], Euler-Bernoulli beam model has been replaced with Timoshenko model and a boundary controller has been proposed to accomplish control goals. However, to guarantee exponential stability of the system the frequency-domain method has been utilized. Recently, in [12]–[15], traditional control objectives like force and vibration control of both types of beam models have been examined. The novelty of these works in contradistinction to previous ones is controlling the orientation of the grasped object in addition to common goals. To prove the stability of the system under designed boundary controller, mathematically complex methods, semigroup theory and frequency domain method, have been employed.

Mahdi Hejrati is M.Sc. Graduate from the School of Mechanical Engineering, Sharif University of Technology, Tehran 11155-9567 Iran (mi.hejrati@gmail.com).

All above-mentioned papers just focused on complicated mathematical tools, e.g., Hilbert Spaces and frequency spectrum analysis, to reach strong stability under simple control. Employing these tools extremely requires expert knowledge, and, moreover, necessitate solving differential equations to reach the stability result. Practically, when one intends to examine the nonlinear dynamics, these approaches become highly inefficient due to needs for solving highly nonlinear differential equations. In contrast, Lyapunov-based procedure is widely well-known and systematic platform that requires no differential equation solving, attracting attentions and paving a way for further examination of these systems. In [5], a Lyapunov-based stability analysis procedure has been proposed. Followingly, this work presents a much-sophisticated stability procedure to reach uniform ultimate boundedness (UUB) stability.

In order to simplify the analysis of the entire system and avoid PDIE (Partial Differential Integral Equation), an alternative is to use a linear dynamic model. However, this simplification may cause model uncertainty that can affect controller performance adversely. From mathematical perspective, an efficient solution is having both simplified, linear model and handling uncertainty. One way to overcome such uncertainty is applying Radial Basic Function Neural-Networks (RBFNNs). Because of their universal approximation ability, RBFNNs are a powerful tool to tackle uncertainty[16]–[21]. Also, having many applications in various eras, flexible arms are exposed to manipulate objects with mass uncertainty or being derived by motors with unknown motor inertia, which implies a significant need for a robust controller, like an adaptive approach. Besides, in many real-life applications, performance of the designed controller can be negatively impacted by input saturation or other input constraints. In [22], a model based controller has been designed to handle a nonlinear backlash along with other control objectives in single link flexible manipulator. In [2], a combined input saturation and dead-zones has been considered in the analysis of single link flexible manipulator. However, none of the above-mentioned issues have been never addressed in the field of dual one-link flexible arms. Considering unknown uncertainties and/or input constraints by the methodology of the previous works will makes the stability analysis complicated and requires further advance research. In this paper, thanks to the proposed stability procedure, not only did we consider combined input constraint, but we also assumed that parameters of constraints are unknown.

Present article, for the first time in the context of dual one-link flexible manipulators, accumulated all the problems mentioned earlier to analyze a more sophisticated model. In contradiction to previous works, we exploited well-known Lyapunov direct method along with energy multiplier method and proposed new procedure to examine the stability, in a way that considering any additional real-world challenges in the model can be handled. For this aim, we split the analysis into two parts: object-position loop and posture loop control. By doing so, we first replaced the complex, direct force control problem with position control approach, and then, applied the backstepping-sliding mode approach to drive all update laws and control signals. It worth to mention that for the first time the backstepping and sliding mode approaches were utilized in dual one-link flexible manipulators control problem that required some modification in the select of sliding surface. The consequence of all this procedure has resulted in a Neuro-Adaptive Boundary Force Control (NABFC) in the presence of unknown model and parameter uncertainty as well as combined unknown input constraints. Compared to the previous publications, the primary contributions of this work are as the following.

1) For the aim of reducing the hardships of identifying parameters in real-world applications, input constraints are assumed to be unknown. Therefore, to handle this and unknown model uncertainty, RBFNNs is employed. Moreover, input saturation and input dead zones are combined to support wide range of constraints.

2) We have developed a strong controller to manipulate any object with unknown mass by using motors with unknown inertia. Combining the backstepping and sliding mode method gives us a more systematic and robust control and update laws. However, introducing such controller to dual one-link flexible manipulators requires overcoming some mathematical challenges.

3) Proposed stability procedure removes requirement of expert knowledge and differential equation solving to reach strong stability, which were necessities of utilized approaches in the previous works. Moreover, presented procedure enables researcher to consider real-world challenges in the model, whereas approaches in the previous works are limited and requires further research and development to be able to handle them in stability analysis.

## II. Dynamic Equations & Preliminaries

Fig. 1 demonstrates the scheme of dual one-link flexible arms. As shown, this system embodies two Euler-Bernoulli arms that are in touch with the grasped object. Each arm is clamped in the root and, at the other side, attached to point mass, which contacts the grasped object. Entire system operates in horizontal plane, and so the gravitational effects are ignored. Also, because the whole dynamic equations of the system are supposed to be linear, all motions in the system are considered to be small. Therefore, the contact forces between concentrated masses and the grasped object are just in shown direction. Table I illustrates parameters of dual one-link flexible manipulator.

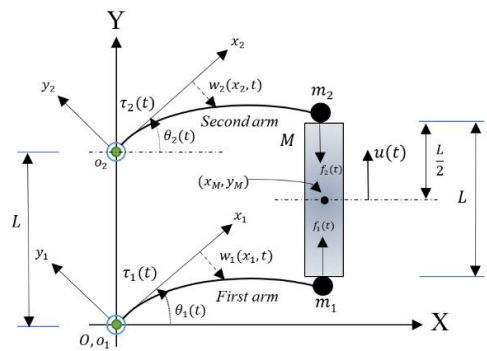

Fig. 1. Schematic drawing of the dual one-link flexible arms

As long as there is a contact between end point masses and grasped object, geometric constraints are persisting. These geometric constraints can be written as below[23],

Table I. Parameters of dual one-link flexible manipulator

| Symbol | Description |
|---|---|
| $m_i$ | Point masses attached to end-effector i |
| $l$ | Length of the flexible arm |
| $M$ | Mass of the object |
| $w_i(x_i, t)$ | Elastic deflection of link i |
| $\theta_i(t)$ | Angular displacement of link i |
| $y_M(t)$ | Vertical displacement of the object |
| $u(t)$ | Control signal of the end-effector |
| $\tau_i(t)$ | Root-applied control signal of link i |
| $\lambda_i(t)$ | Lagrange multiplier |
| $EI_i$ | Uniform flexural density of link i |
| $\rho_i$ | Uniform mass per unit length of link i |
| $J_i$ | Inertia moment of the rotor hub i |
| $\Delta F, \Delta f_i, i=1,2$ | Model Uncertainties |

$$\phi_i(t) = l\theta_i(t) - w_i(l,t) - y_M + \frac{L}{2} = 0 \quad i = 1,2. \quad (1)$$

This equation does own a significant importance because it reduces the complex force control task to a well-known position control yet ensures controlling contact force without force sensors. However, in order to drive hybrid PDE-ODE dynamic equations the Hamilton's principle is employed. For this purpose, the Kinetic and Potential energy of entire system are required which can be expressed as $KE = \sum_{i=1}^{2}\{\frac{\rho_i}{2}\int_0^l[x_i\dot{\theta}_i(t) - \dot{w}_i(x_i,t)]^2 dx_i + \frac{m_i}{2}[l\dot{\theta}_i(t) - \dot{w}_i(l,t)]^2 + \frac{J_i}{2}\dot{\theta}^2(t)\} + \frac{M}{2}\dot{y}_M^2(t)$ and $PE = \sum_{i=1}^{2}\frac{EI_i}{2}\int_0^l[w_{xxi}(x_i,t)]^2 dx_i$, where a dot denotes the time derivatives and a $\frac{\partial}{\partial x}$ denotes the partial derivatives with respect to the spatial variable. Moreover, the virtual work consisting of two torques at the root and one at the end-effector can be written as $W = \sum_{i=1}^{2}\tau_i(t)\delta\theta_i(t) + u\delta y_M$. Therefore, by applying Hamilton's principle and Lagrange multiplier method and considering model uncertainty, the resulting equations of motion can be derived as below,

$$\ddot{w}_i(x_i,t) + \frac{EI_i}{\rho_i}w_{xxxxi}(x_i,t) = x_i\ddot{\theta}_i(t) \quad (2)$$

$$w_i(0,t) = w_{xi}(0,t) = w_{xxi}(l,t) = 0 \quad (3)$$

$$m_i\{\ddot{w}_i(l,t) - l\ddot{\theta}_i(t)\} - EI_iw_{xxxi}(l,t) + \lambda_i(t) = 0 \quad (4)$$

$$J_i\ddot{\theta}_i(t) + EI_iw_{xxi}(0,t) = \tau_i + \Delta f_i \quad (5)$$

$$m\ddot{y}_M(t) + \{\lambda_1 + \lambda_2\} = u + \Delta F \quad (6)$$

$$\phi_i(t) = l\theta_i(t) - w_i(l,t) - y_M + \frac{L}{2} = 0 \quad i = 1,2 \quad (7)$$

where all parameters are defined in Table I. In order to circumvent nonhomogeneous boundary condition and eliminate $\lambda_i(t), i = 1,2$, by combining (6) and (3) through constraint equation (7), we get to,

$$\ddot{w}_i(x_i,t) + \frac{EI_i}{\rho_i}w_{xxxxi}(x_i,t) = x_i\ddot{\theta}_i(t) \quad (8)$$

$$w_i(0,t) = w_{xi}(0,t) = w_{xxi}(l,t) = 0 \quad (9)$$

$$J_i\ddot{\theta}_i(t) + EI_iw_{xxi}(0,t) = \tau_i + \Delta f_i \quad (10)$$

$$m\ddot{y}_M(t) + \sum_{i=1}^{2}EI_iw_{xxxi}(l,t) = u + \Delta F \quad (11)$$

$$\phi_i(t) = l\theta_i(t) - w_i(l,t) - y_M + \frac{L}{2} = 0 \quad i = 1,2 \quad (12)$$

and the related algebraic relationships as below,

$$m\lambda_1(t) = m_1(u + \Delta F) + (m_2 + M)EI_1w_{xxx1}(l,t) - m_1EI_2w_{xxx2}(l,t) \quad (13)$$

$$m\lambda_2(t) = m_2(u + \Delta F) + (m_1 + M)EI_2w_{xxx2}(l,t) - m_2EI_1w_{xxx1}(l,t) \quad (14)$$

where the relationship between $\lambda_i(t)$ and $f_i(t), i = 1,2$, the reaction forces, is as $f_i(t) = (-1)^i\lambda_i(t)$.

At the following, some useful lemmas and inequalities are presented.

*Assumption 1*: Model uncertainties $\Delta F$ and $\Delta f_i, i = 1,2$ are supposed to have unknown positive constants $\varpi$ and $\kappa_i$ that satisfy $|\Delta F| \leq \varpi$ and $|\Delta f_i| \leq \kappa_i$. Besides, these uncertainties contain the characteristic of $\lim_{t\to\infty}\Delta F(t) = \lim_{t\to\infty}\Delta f_i(t) = 0$.

*Lemma 1*[24]: For functions of $\phi_1(x,t), \phi_2(x,t) \in \mathbb{R}$, the following inequality hold,

$$\phi_1(x,t)\phi_2(x,t) \leq \frac{1}{2}\{\phi_1^2(x,t) + \phi_2^2(x,t)\} \leq \omega_1\phi_1^2(x,t) + \frac{1}{\omega_1}\phi_2^2(x,t).$$

where $\omega_1 > 0$ is a constant. Furthermore, for a continuous differentiable function $\Phi(x,t)$ with respect to $x$, under the condition of $\Phi(0,t) = 0$ and $\Phi_x(0,t) = 0$, following inequalities hold,

$$\Phi^2(x,t) \leq l\int_0^l \Phi_x^2(x,t)dx$$

$$\Phi_x^2(x,t) \leq l\int_0^l \Phi_{xx}^2(x,t)dx$$

also, the following one is ascertainable,

$$\Phi^2(x,t) \leq l^3\int_0^l \Phi_{xx}^2(x,t)dx.$$

*Lemma 2*[24]: Th following inequality hold for $\varrho(t) \in \mathbb{R}$,

$$|\varrho(t)| - \varrho(t)\tanh\left(\frac{\varrho(t)}{\omega_2}\right) \leq 0.2785\omega_2$$

where $\omega_2 > 0$ is a constant.

*Lemma 3*[25]: Any arbitrary unknown continuous function $\mathcal{H}(h): \mathbb{R}^m \to \mathbb{R}$, can be approximated by the means of RBF neural networks as below,

$$\mathcal{H}(h) = \widehat{W}^T\Psi(h) + \varepsilon$$

where $h = [h_1, h_2, \cdots, h_n]^T \in \mathbb{R}^n$, $\Psi(h) = [\Psi_1(h), \Psi_2(h), \cdots, \Psi_n(h)]^T \in \mathbb{R}^n$, $\widehat{W} \in \mathbb{R}^n$, and $\varepsilon$ are input, basic function, weight vector, and approximation bias of the neural networks, respectively. The optimal weight vector $W^*$ can be expressed by,

$$W^* = arg \min_{\widehat{W} \in F_s} \left\{ \sup_{h \in F_r} |\widehat{\mathcal{H}}(h|\widehat{W}) - \mathcal{H}(h)| \right\}$$

where $F_s = \{\widehat{W} | \|\widehat{W}\| \leq \omega_3\}$ is valid field of the vector with a design value $\omega_3$, allowable set of the state vector $F_r$, and $\widehat{\mathcal{H}}(h|\widehat{W}) = \widehat{W}^T \Psi(h)$. Therefore, we can write,

$$\mathcal{H}(h) = W^{*T} \Psi(h) + \varepsilon^*$$

with $\varepsilon^*$ is the ideal approximation bias that satisfies $|\varepsilon^*| \leq \omega_4$, where $\omega_4$ is an unknown positive constant. Moreover, the bias function is Gaussian function described in [26].

This paper aims to control the contact force between the end-effector and the grasped object in the presence of unknown combined input constraints and unknown model and parameter uncertainties. However, having a tool to reduce the force control task to a position control will be exceedingly beneficial, and dual one-link flexible manipulators, fortunately, own such an implement: geometric constraints. Nevertheless, we need to derive control objectives that successfully relate force and position control approaches in order to have such a reduction.

Thus, the designed controller, at the static position, must be able to satisfy $\lambda_i \to \lambda_i^d$, $\dot{w}_i(x_i, t) \to 0$, $\dot{\theta} \to 0$, and $\dot{y}_M \to 0$. Therefore, in this desired position we have,

$$\lambda_i = \lambda_i^d, \dot{w}_i(x_i, t) = 0, \dot{\theta} = 0, \dot{y}_M = 0, \quad (15)$$

where $i = 1,2$. So, by substituting (15) in (8-11) and using (13) and (14) and Assumption 1, we can get to desired static position as,

$$\lambda_1^d = -\lambda_2^d, \quad w_i^d(x_i) = \frac{\lambda_i^d x_i^2}{2EI_i} \left( \frac{x_i}{3} - l \right),$$

$$\theta_1^d = \theta_2^d - \frac{\lambda_1^d l^2}{3} \left( \frac{EI_1 + EI_2}{EI_1 EI_2} \right), \quad (16)$$

$$y_M^d = l\theta_1^d + \frac{l^3 \lambda_1^d}{3EI_1}.$$

Due to geometric constraints, it is clear from (16) that $\lambda_1^d$ and $\lambda_2^d$ are interdependent and cannot be chosen separately. Furthermore, $\theta_1^d$ and $\theta_2^d$ have relationship through $\lambda_1^d$, which means just by choosing two desired values remains will be derived automatically. Besides, because of the inherent characteristics of dual one-link flexible manipulators resulted from (16), just by controlling angular positions of rotors alongside the position of the grasped object, we can accomplish the grasping task without the need for force sensor. Finally, our control goals are as below,

$$y_M \to y_M^d, \theta_i(t) \to \theta_i^d, \dot{y}_M \to 0, \dot{\theta}_i \to 0. \quad (17)$$

In many real-life applications, actuators are unable to provide all of the required power supplies and virtually confront some issues like input saturation and input dead zones. These limitations can cause the controller to operate poorly, resulting in failure or other undesirable outcomes. Hence, not only the input constraint, but also the combination of input saturation and input dead zones is considered in this paper.

*Lemma 5*[27]: If $\Sigma$ is supposed to be a signal exerted to the system and $\varsigma$ be the designed control signal, the issue of input saturation along with input dead zones can be transformed into an equivalent input saturation form as,

$$\Sigma = D\left(sat(D^+(\varsigma))\right)$$
$$= \begin{cases} k_r(k_M - m_r) & if \quad \varsigma \geq k_r(k_M - m_r) \\ \varsigma & if \quad k_l(k_m - m_l) < \varsigma < k_r(k_M - m_r) \\ k_l(k_m - m_l) & if \quad \varsigma \leq k_l(k_m - m_l). \end{cases} \quad (18)$$

Considering Lemma 5, we can define actuators' output as,

$$\tau_i = D\left(sat(D^+(\bar{\tau}_i))\right)$$
$$= \begin{cases} k_{r1}(k_{M1} - m_{r1}) & if \quad \bar{\tau}_i \geq k_r(k_{M1} - m_{r1}) \\ \bar{\tau}_i & if \quad k_{l1}(k_{m1} - m_{l1}) < \bar{\tau}_i < k_{r1}(k_{M1} - m_{r1}) \\ k_{l1}(k_{m1} - m_{l1}) & if \quad \bar{\tau}_i \leq k_{l1}(k_{m1} - m_{l1}), \end{cases} \quad (19)$$

and

$$u = D\left(sat(D^+(v))\right)$$
$$= \begin{cases} k_{r2}(k_{M2} - m_{r2}) & if \quad v \geq k_{r2}(k_{M2} - m_{r2}) \\ v & if \quad k_{l2}(k_{m2} - m_{l2}) < v < k_{r2}(k_{M2} - m_{r2}) \\ k_{l2}(k_{m2} - m_{l2}) & if \quad v \leq k_l(k_{m2} - m_{l2}), \end{cases} \quad (20)$$

where $k_{Mi} > 0$ $and$ $k_{mi} < 0$ are unknown upper and lower saturation levels, $m_{li} < 0$ $and$ $m_{ri} > 0$ are the unknown dead zones ranges, and $k_{li} > 0$ $and$ $k_{ri} < 0$ are the unknown slope parameters, $i = 1,2$.

In order to have robust controller with ability of cancelling unknown input constraints and unknown uncertainties, let $\Delta \tau_i = \tau_i - \bar{\tau}_i$ and $\Delta u = u - v$, and by defining new unknown variable $\Delta_i = \Delta \tau_i + \Delta f_i$ and $\bar{\Delta} = \Delta F + \Delta u$, we can rewrite (10) and (11) as below,

$$J_i \ddot{\theta}_i(t) + EI_i w_{xxi}(0,t) = \bar{\tau}_i + \Delta_i \quad (21)$$

$$m\ddot{y}_M(t) + \sum_{i=1}^{2} EI_i w_{xxxi}(l,t) = v + \bar{\Delta}. \quad (22)$$

To simplify the remaining procedure, we move the equilibrium points to the origin. Letting $e_i(t) = \theta_i(t) - \theta_i^d$, $q_i(x_i, t) = w_i(x_i, t) - w_i^d(x_i)$, $p(t) = y_M(t) - y_M^d$, governing equations can be transformed to the following version,

$$\ddot{q}_i(x_i, t) + \frac{EI_i}{\rho_i} q_{xxxxi}(x_i, t) = x_i \ddot{e}_i(t) \quad (23)$$

$$q_i(0,t) = q_{xi}(0,t) = q_{xxi}(l,t) = 0 \quad (24)$$

$$m\ddot{p}(t) + \sum_{i=1}^{2} EI_i q_{xxxi}(l,t) = v + \bar{\Delta} \quad (25)$$

$$J_i \ddot{e}_i(t) + EI_i q_{xxi}(0,t) = \bar{\tau}_i + \Delta_i + \lambda_i^d l. \quad (26)$$

## III. Controller Designing & Stability Analysis

In this section we focused on deriving proper control laws and updating rules in order to stabilize the system. In order to derive NABFC laws, we adopted a novel, integrated Backstepping-Sliding mode method, not only to have a systematic procedure but also to achieve a more robust controller. Literally, both the backstepping and sliding mode control methods are thoroughly studied. However, utilizing these methods in the context of dual one-link flexible manipulators happens for the first time in this paper. By the way, the procedure is divided into two distinct steps: object-position loop and posture loop. After applying Lyapunov's stability procedure to each step, the general Lyapunov function is purposefully defined to assemble all parts and finish the analysis.

To continue, suppose $r_1(t) = p(t), r_2(t) = \dot{p}(t), r_{3i}(t) = e_i(t)$, and $r_{4i}(t) = \dot{e}_i(t)$. Then, (25) and (26) can be rewritten as follow,

$$\dot{r}_1(t) = r_2(t) \tag{27}$$

$$\dot{r}_2(t) = \frac{1}{m}\left\{v + \bar{\Delta} - \sum_{i=1}^{2} EI_i q_{xxxi}(l,t)\right\} \tag{28}$$

$$\dot{r}_{3i}(t) = r_{4i}(t) \tag{29}$$

$$\dot{r}_{4i}(t) = \frac{1}{J_i}\left\{\bar{\tau}_i + \Delta_i + \lambda_i^d l - EI_i q_{xxi}(0,t)\right\}. \tag{30}$$

### A. Object-Position Loop Control Design

Dividing controller design process into separate parts can facilitate the procedure and avoid complexity. In this part, the end-effector control signal is derived to cancel some part of model uncertainties, unknown grasped-object mass, and unknown input constraints associated with this signal.

Let define,

$$z_1(t) = r_1(t) \tag{31}$$

$$z_2(t) = r_2(t) - \alpha_1(t) \tag{32}$$

$$s(t) = \bar{\mu} z_1(t) + z_2(t) \tag{33}$$

where $z_1, z_2$ are stabilizing functions, $s$ is sliding surface, $\bar{\mu}$ is constant parameter, and $\alpha_1(t)$ is virtual control, which can be defined as below,

$$\alpha_1(t) = -c_1 z_1(t) - \sum_{i=1}^{2} q_{xxxi}(l,t), \tag{34}$$

with $c_1 > 0$ being a constant parameter. The derivative of (34) can be computed as below,

$$\dot{\alpha}_1(t) = -c_1 \dot{z}_1(t) - \sum_{i=1}^{2} \dot{q}_{xxxi}(l,t).$$

However, as we mentioned in Lemma 3, the unknown uncertain term, $\bar{\Delta}$, can be estimated by RBFNN as below,

$$\bar{\Delta} = W^{*T} Y(X) + \varepsilon^* \tag{35}$$

where $W^*$ and $\varepsilon^*$ are the unknown ideal weight vector and optimal approximation error, respectively, $Y(X)$ is the basis function, and $X = [z_1(t), s(t), \dot{\alpha}_1(t), w_{xxx1}(l,t), w_{xxx2}(l,t), \dot{w}_{xxx1}(l,t), \dot{w}_{xxx2}(l,t)]^T$. In addition, from Lemma 3, there is an unknown constant $\bar{\varepsilon} > 0$ such that $|\varepsilon^*| \leq \bar{\varepsilon}$.

**Theorem 1.** Suppose a system with governing equations of motion like (27) and (28) that contains unknown uncertainty $\bar{\Delta}$ and unknown parameter $m$. Handling mentioned issues, control signal $v$ can be proposed as,

$$v(t) = -\eta \, sign(s) - ks - z_1 - \widehat{m} \dot{z}_1 (\bar{\mu} + c_1)$$
$$+ \sum_{i=1}^{2} EI_i q_{xxxi}(l,t) - \widehat{W}^T Y(X) - \hat{\bar{\varepsilon}} \tanh \frac{s}{\epsilon_1} \tag{36}$$
$$- \widehat{m} \sum_{i=1}^{2} \dot{q}_{xxxi}(l,t)$$

with the update laws as below,

$$\dot{\widetilde{W}} = -\dot{\widehat{W}} = -a_1 s Y(X) + a_1 a_2 \widehat{W} \tag{37}$$

$$\dot{\tilde{\bar{\varepsilon}}} = -\dot{\hat{\bar{\varepsilon}}} = -\gamma_1 s \tanh \frac{s}{\epsilon_1} + \gamma_1 \gamma_2 \hat{\bar{\varepsilon}} \tag{38}$$

$$\dot{\widetilde{m}} = -\dot{\widehat{m}}$$
$$= -b_1 s \dot{z}_1 (\bar{\mu} + c_1) + b_1 b_2 \widehat{m} \tag{39}$$
$$- b_1 s \sum_{i=1}^{2} \dot{q}_{xxxi}(l,t)$$

with $\eta, k, \bar{\mu}, c_1$, and $\epsilon_1$ being positive and constant parameter. Also, $\widehat{m}$ is estimation of $m$, $\widehat{W}$ is estimation of $W^*$, and $\hat{\bar{\varepsilon}}$ is estimation of $\bar{\varepsilon}$. Employing (36) leads to,

$$\dot{V}_o(t) \leq -z_1^2(t)\{(\bar{\mu} + c_1) - \vartheta_1\} - \eta|s| - ks^2$$
$$+ \frac{1}{\vartheta_1}\left\{\sum_{i=1}^{2} q_{xxxi}(l,t)\right\}^2 - b_2 \widetilde{m}^2 \left(1 - \frac{1}{b_3}\right)$$
$$- \gamma_2 \hat{\bar{\varepsilon}}^2 \left(1 - \frac{1}{\gamma_3}\right) - \frac{1}{2} a_2 \widetilde{W}^T \widetilde{W} + \frac{1}{2} a_2 \overline{W}^{*T} \overline{W}^* \tag{40}$$
$$+ \gamma_3 \gamma_2 \bar{\varepsilon}^2 + b_2 b_3 m^2 + 0.2785 \epsilon_1 \bar{\varepsilon}$$

with $a_1, a_2, b_1, b_2, b_3, \gamma_2, \gamma_3, \vartheta_1$ being positive and constant parameter.
**Proof.**

Let $V_1(t)$ as,

$$V_1(t) = \frac{1}{2} z_1^2(t) \tag{41}$$

and by using (27), (31), (32), and (34), time derivative of (41) can be computed as,

$$\dot{V}_1(t) = -c_1 z_1^2(t) + z_1(t) z_2(t) - z_1(t) \sum_{i=1}^{2} q_{xxxi}(l,t). \tag{42}$$

Substituting (32) in (33), taking time derivative, and using (28) lead to,

$$\dot{s}(t) = (\bar{\mu} + c_1)\dot{z}_1(t) + \frac{1}{m}\left\{v + \bar{\Delta} - \sum_{i=1}^{2} EI_i q_{xxxi}(l,t)\right\} \\ + \sum_{i=1}^{2} \dot{q}_{xxxi}(l,t). \tag{43}$$

Selecting second Lyapunov function as,

$$V_2(t) = V_1(t) + \frac{1}{2}ms^2(t) \tag{44}$$

and calculating its time derivative by using (33), (42) and (43), we get to,

$$\dot{V}_2(t) = -(\bar{\mu} + c_1)z_1^2(t) - z_1(t)\sum_{i=1}^{2} q_{xxxi}(l,t) \\ + s\left\{m(\bar{\mu}+c_1)\dot{z}_1(t) + z_1 + v + \bar{\Delta} \\ - \sum_{i=1}^{2} EI_i q_{xxxi}(l,t) + m\sum_{i=1}^{2} \dot{q}_{xxxi}(l,t)\right\}. \tag{45}$$

Therefore, by substituting (35) and (36) in (45) and calling Lemma 2 alongside $|\varepsilon^*| \leq \bar{\varepsilon}$,

$$\dot{V}_2(t) \leq -(\bar{\mu} + c_1)z_1^2(t) - z_1(t)\sum_{i=1}^{2} q_{xxxi}(l,t) \\ + \tilde{m}s\left\{(\bar{\mu}+c_1)\dot{z}_1(t) + \sum_{i=1}^{2}\dot{q}_{xxxi}(l,t)\right\} \\ + s\tilde{W}^T Y(X) - \eta|s| - ks^2 + s\tilde{\varepsilon}\tanh\frac{s}{\epsilon_1} \\ + 0.2785\bar{\varepsilon}\epsilon_1 \tag{46}$$

where $\tilde{W} = W^* - \hat{W}, \tilde{\varepsilon} = \bar{\varepsilon} - \hat{\varepsilon}, \tilde{m} = m - \hat{m}$.

Final Lyapunov function of this part can be propose as below,

$$V_0(t) = V_2(t) + \frac{1}{2a_1}\tilde{W}^T\tilde{W} + \frac{1}{2\gamma_1}\tilde{\varepsilon}^2 + \frac{1}{2b_1}\tilde{m}^2. \tag{47}$$

Invoking (37), (38), (39), and (46) along with Lemma 3, time derivative of (47) can be calculated as follow,

$$\dot{V}_o(t) \leq -z_1^2(t)\{(\bar{\mu}+c_1) - \vartheta_1\} - \eta|s| - ks^2 \\ + \frac{1}{\vartheta_1}\left\{\sum_{i=1}^{2} q_{xxxi}(l,t)\right\}^2 - b_2\tilde{m}^2\left(1-\frac{1}{b_3}\right) \\ - \gamma_2\tilde{\varepsilon}^2\left(1-\frac{1}{\gamma_3}\right) - \frac{1}{2}a_2\tilde{W}^T\tilde{W} + \frac{1}{2}a_2 W^{*T}W^* \\ + \gamma_3\gamma_2\bar{\varepsilon}^2 + b_2 b_3 m^2 + 0.2785\epsilon_1\bar{\varepsilon}. \tag{48}$$

B. *Posture Loop Control Design*

In this section, both control signal of applying desired force to the object through root actuators or update rules to overcome unknown uncertainties are drawn. In contrast to the to-end-effector-applied control signal, which helps to have more stable states and vibration removal, posture control signals play a significant role in the implementation of desired force.

Let define,

$$z_{3i}(t) = r_{3i}(t) \tag{49}$$

$$z_{4i}(t) = r_{4i}(t) - \alpha_{2i}(t) \tag{50}$$

$$s_i(t) = \mu_i z_{3i}(t) + z_{4i}(t) \tag{51}$$

where $z_{3i}, z_{4i}$ are stabilizing functions, $s_i$ is sliding surface, $\mu_i$ is constant parameter, and $\alpha_{2i}(t)$ is virtual control, $i = 1,2$, which can be defined as below,

$$\alpha_{2i}(t) = -c_{3i}z_{3i}(t), \tag{52}$$

with $c_{3i} > 0$ being constant parameter. Taking time derivative of (52) leads to,

$$\dot{\alpha}_{2i}(t) = -c_{3i}\dot{z}_{3i}(t).$$

However, as we mentioned in Lemma 3, the unknown uncertain term, $\Delta_i$, can be estimated by RBFNN as below,

$$\Delta_i = U_i^{*T}Y(Z_i) + \pi_i^* \tag{53}$$

where $U_i^*$ and $\pi_i^*$ are the unknown ideal weight vector and optimal approximation error, respectively, $Y(Z)$ is the basis function, and $Z_i = [z_{3i}(t), s_i(t), \dot{\alpha}_{2i}(t)]^T$. In addition, from Lemma 3, there is an unknown constant $\bar{\pi}_i > 0$ such that $|\pi_i^*| \leq \bar{\pi}_i$.

*Theorem 2.* Suppose a system with governing equations of motion like (29) and (30) that contains unknown uncertainty $\Delta_i$ and unknown parameters $J_i$. Handling mentioned issues, control signal $\bar{\tau}_i$ can be proposed as,

$$\bar{\tau}_i(t) = -\xi_i sign(s_i) - k_i s - z_{3i} - \hat{J}_i \dot{z}_{3i}(\mu_i + c_{3i}) \\ + EI_i q_{xxi}(0,t) - \hat{U}_i^T Y(Z_i) - \hat{\bar{\pi}}_i \tanh\frac{s_i}{\epsilon_{2i}} \\ - \lambda_i^d l, \tag{54}$$

with update rules of unknown parameters as,

$$\dot{\tilde{U}}_i = -\dot{\hat{U}}_i = -a_{3i}s_i Y(Z_i) + a_{3i}a_{4i}\hat{U}_i \tag{55}$$

$$\dot{\tilde{\bar{\pi}}}_i = -\dot{\hat{\bar{\pi}}}_i = -\zeta_{1i}s_i \tanh\frac{s_i}{\epsilon_{2i}} + \zeta_{1i}\zeta_{2i}\hat{\bar{\pi}} \tag{56}$$

$$\dot{\tilde{J}}_i = -\dot{\hat{J}}_i = -g_{1i}s_i\dot{z}_{3i}(\mu_i+c_{3i}) + g_{1i}g_{2i}\hat{J}_i \tag{57}$$

with $\xi_i, k_i, \mu_i, c_{3i}$, and $\epsilon_{2i}$ being positive and constant parameters. Additionally, $\hat{J}_i$ are estimation of $J$, $\hat{U}_i$ is estimation of $U_i^*$, and $\hat{\bar{\pi}}_i$ is estimation of $\bar{\pi}_i$. Exploiting (54)-(57) leads to,

$$\dot{V}_{Ii}(t) \leq -(\mu_i + c_{3i})z_{3i}^2 - \xi_i|s_i| - k_i s_i^2 \\ - g_{2i}\tilde{J}_i^2\left(1-\frac{1}{g_{3i}}\right) - \zeta_{2i}\tilde{\bar{\pi}}_i^2\left(1-\frac{1}{\zeta_{3i}}\right) \\ - \frac{1}{2}a_{4i}\tilde{U}_i^T\tilde{U}_i + \frac{1}{2}a_{4i}U_i^{*T}U_i^* + \zeta_{2i}\zeta_{3i}\bar{\pi}_i^2 \\ + g_{2i}g_{3i}J_i^2 + 0.2785\epsilon_{2i}\bar{\pi}_i \tag{58}$$

with $a_{3i}, a_{4i}, g_{1i}, g_{2i}, g_{3i}, \zeta_{1i}, \zeta_{2i}, \zeta_{3i}$ being positive and constant parameter.

*Proof.*

Let $V_{3i}(t)$ as,

$$V_{3i}(t) = \frac{1}{2}z_{3i}^2(t) \quad (59)$$

and by using (27), (49), (50), and (52), time derivative of (59) can be computed as,

$$\dot{V}_{3i}(t) = -c_{3i}z_{3i}^2(t) + z_{3i}(t)z_{3i}(t). \quad (60)$$

Substituting (50) in (51), taking time derivative, and making use of (30) lead to,

$$\dot{s}_i(t) = (\mu_i + c_{3i})\dot{z}_{3i}(t) + \frac{1}{J_i}\{\bar{\tau}_i + \Delta_i + \lambda_i^d l - EI_i q_{xxi}(0,t)\}. \quad (61)$$

Selecting second Lyapunov function as,

$$V_{4i}(t) = V_{3i}(t) + \frac{1}{2}J_i s_i^2(t) \quad (62)$$

and calculating its time derivative by using (51), (60) and (61), we achieve,

$$\dot{V}_{4i}(t) = -(\mu_i + c_{3i})z_{3i}^2(t) + s_i\{J_i(\mu_i + c_{3i})\dot{z}_{3i}(t) + z_{3i} + \bar{\tau}_i + \Delta_i + \lambda_i^d l - EI_i q_{xxi}(0,t)\}. \quad (63)$$

Therefore, by substituting (53) and (54) in (63) and calling Lemma 2 in conjunction with $|\pi_i^*| \leq \bar{\pi}_i$, we get to,

$$\dot{V}_{4i}(t) \leq -(\mu_i + c_{3i})z_{3i}^2(t) + \tilde{J}_i s_i(\mu_i + c_{3i})\dot{z}_{3i}(t) + s_i \tilde{U}_i^T Y(Z_i) - \zeta_i|s_i| - k_i s_i^2 + s_i \tilde{\bar{\pi}}_i \tanh\frac{s_i}{\epsilon_{2i}} + 0.2785\bar{\pi}_i\epsilon_{2i} \quad (64)$$

where $\tilde{U} = U^* - \hat{U}, \tilde{\bar{\pi}} = \bar{\pi} - \hat{\bar{\pi}}, \tilde{J} = J - \hat{J}$.

Final Lyapunov function of this part can be proposed as below,

$$V_{li}(t) = V_{4i}(t) + \frac{1}{2a_{3i}}\tilde{U}_i^T \tilde{U}_i + \frac{1}{2\zeta_{1i}}\tilde{\bar{\pi}}_i^2 + \frac{1}{2g_{1i}}\tilde{J}_i^2. \quad (65)$$

Invoking (55), (56), (57), and (64) along with Lemma 3, time derivative of (65) can be calculated as follow,

$$\dot{V}_{li}(t) \leq -(\mu_i + c_{3i})z_{3i}^2 - \xi_i|s_i| - k_i s_i^2 \\ - g_{2i}\tilde{J}_i^2\left(1 - \frac{1}{g_{3i}}\right) - \zeta_{2i}\tilde{\bar{\pi}}_i^2\left(1 - \frac{1}{\zeta_{3i}}\right) \\ - \frac{1}{2}a_{4i}\tilde{U}_i^T \tilde{U}_i + \frac{1}{2}a_{4i}U_i^{*T}U_i^* + \zeta_{2i}\zeta_{3i}\bar{\pi}_i^2 \\ + g_{2i}g_{3i}J_i^2 + 0.2785\epsilon_{2i}\bar{\pi}_i. \quad (66)$$

### C. Convergence Analysis

In this part, the closed-loop stability of the whole system based on the control and update laws represented by (36)-(39) and (54)-(57) is established.

***Theorem 3.*** Suppose the dual one-link flexible manipulator's representative system equation is (23)-(30). Based on the NABFC laws described by (36)-(39) and (54)-(57), the closed-loop system of the dual one-link flexible manipulator is uniformly ultimately bounded.

***Proof.*** Lyapunov function can be constructed as,

$$H(t) = V_o(t) + E(t) + C(t) + \sum_{i=1}^{2}V_{li}(t) \quad (67)$$

where,

$$E(t) = \frac{\beta_1}{2}\sum_{i=1}^{2}\{EI_i\int_0^l q_{xxi}^2(x,t)dx_i + \rho_i\int_0^l \dot{y}_i^2(x_i,t)dx_i\} \quad (68)$$

$$C(t) = \beta_2\sum_{i=1}^{2}\rho_i\int_0^l (x_i - l)\dot{y}_i(x_i,t)q_{xi}(x_i,t)dx_i, \quad (69)$$

$\beta_1$ and $\beta_2$ are positive constants, and $y_i(x_i,t) = q_i(x_i,t) - x_i e_i(t)$. To guarantee that (67) is a positive Lyapunov function and, also, to have a well-sequenced stability analysis, the procedure is divided into two steps. Firstly, positiveness of the $H(t)$ is established, and, afterward, closed-loop stability is proved.

*Step 1*: Considering (69) and employing Cauchy-Schwarz inequality with Lemma 1, it derives,

$$|C(t)| \leq \beta_2\sum_{i=1}^{2}\rho_i\int_0^l |(x_i - l)\dot{y}_i(x_i,t)q_{xi}(x_i,t)|dx_i \\ \leq \beta_2 l\sum_{i=1}^{2}\rho_i\int_0^l \left(\dot{y}_i^2(x_i,t) + q_{xi}^2(x_i,t)\right)dx_i \\ \leq \beta_2 l\sum_{i=1}^{2}\rho_i\int_0^l \left(\dot{y}_i^2(x_i,t) + l^2 q_{xxi}^2(x_i,t)\right)dx_i \\ \leq \varphi_1 E(t) \quad (70)$$

where $\varphi_1 = \frac{\beta_2 l \min(\rho_1,\rho_2)\min(1,l^2)}{0.5\beta_1 \min(EI_1,EI_2,\rho_1,\rho_2)}$. Then we can complete first step by utilizing absolute value property and adding terms in (67), $C(t)$ excluded, to (70), we have,

$$0 \leq \varphi_2\left\{E(t) + V_o(t) + \sum_{i=1}^{2}V_{li}(t)\right\} \leq H(t) \\ \leq \varphi_3\left\{E(t) + V_o(t) + \sum_{i=1}^{2}V_{li}(t)\right\} \quad (71)$$

where, $\varphi_2 = min\{1 - \varphi_1, 1\}, \varphi_3 = max\{1 + \varphi_1, 1\}$. Moreover, $\beta_1$ and $\beta_2$ must choose in a way that positiveness of the $H(t)$ guarantee.

*Step 2*: To proceed, we need to derive time derivative of (67). With the aim of simplicity, we take the time derivative of each term in (67) separately. Consequently, taking time derivate of (68) and using (23) along with boundary conditions, lead to,

$$\dot{E}(t) = \beta_1 \sum_{i=1}^{2} EI_i \left\{ \int_0^l q_{xxi}(x_i,t)\dot{q}_{xxi}(x_i,t)dx_i \right.$$
$$\left. - \int_0^l \dot{y}_i(x_i,t)q_{xxxxi}(x_i,t)dx_i \right\} \quad (72)$$
$$= -\beta_1 \sum_{i=1}^{2} EI_i \{-\dot{y}_i(l,t)q_{xxxi}(l,t)$$
$$- \dot{y}_{xi}(0,t)q_{xxi}(0,t)\}.$$

Now, by comparing $y_i(l,t) = q_i(l,t) - le_i(t)$ and $p(t) = le_1(t) - q_1(l,t) = le_2(t) - q_2(l,t)$, one can conclude that $p(t) = -y_1(l,t) = -y_2(l,t)$, that leads to $-\sum_{i=1}^{2}\dot{y}_i(l,t)q_{xxxi}(l,t) = \dot{p}(t)\sum_{i=1}^{2}q_{xxxi}(l,t)$. Therefore, invoking $\dot{p}(t)\sum_{i=1}^{2}q_{xxxi}(l,t) = -\frac{1}{2}\dot{p}^2(t) - \frac{1}{2}\left(\sum_{i=1}^{2}q_{xxxi}(l,t)\right)^2 + \frac{1}{2}s^2 - \frac{1}{2}(\bar{u}+c_1)z_1^2 - z_1(\bar{u}+c_1)\dot{p}(t) - (\bar{u}+c_1)z_1\sum_{i=1}^{2}q_{xxxi}(l,t)$, the fact that $\dot{y}_{xi}(0,t) = -\dot{e}_i(t)$, $\dot{e}(t) = s_i - (\mu_i + c_{3i})z_{3i}$, and Lemma 1, the time derivative of $E(t)$ can be derived as below,

$$\dot{E}(t) \leq \beta_1 \dot{p}^2(t) \sum_{i=1}^{2} EI_i \left\{\frac{1}{2} - \vartheta_2(\bar{u}+c_1)\right\}$$
$$- \beta_1(\bar{u}+c_1)\left\{\frac{1}{2} - \vartheta_3 - \frac{1}{\vartheta_2}\right\}(EI_1 + EI_2)z_1^2$$
$$- \beta_1(EI_1 + EI_2)\left\{\frac{1}{2}\right.$$
$$\left. - \frac{(\bar{u}+c_1)}{\vartheta_3}\right\}\left(\sum_{i=1}^{2}q_{xxxi}(l,t)\right)^2 \quad (73)$$
$$+ \frac{1}{2}\beta_1(EI_1+EI_2)s^2 + \beta_1 \sum_{i=1}^{2} EI_i\vartheta_{4i}s_i^2$$
$$+ \beta_1 \sum_{i=1}^{2} EI_i(\mu_i + c_{3i})\frac{1}{\vartheta_{5i}}z_{3i}^2$$
$$+ \beta_1 \sum_{i=1}^{2} EI_i q_{xxi}^2(0,t)\left\{\frac{1}{\vartheta_{4i}} + \vartheta_{5i}(\mu_i + c_{3i})\right\}.$$

Considering (69), $\dot{C}(t)$ can be computed as,

$$\dot{C}(t) = \beta_2 \sum_{i=1}^{2}\rho_i \int_0^l (x_i-l)\ddot{y}_i(x_i,t)q_{xi}(x_i,t)dx_i$$
$$+ \beta_2 \sum_{i=1}^{2}\rho_i \int_0^l (x_i-l)\dot{y}_i(x_i,t)\dot{q}_{xi}(x_i,t)dx_i \quad (74)$$
$$= \dot{C}_1(t) + \dot{C}_2(t)$$

where,

$$\dot{C}_1(t) = \beta_2 \sum_{i=1}^{2}\rho_i \int_0^l (x_i-l)\ddot{y}_i(x_i,t)q_{xi}(x_i,t)dx_i \quad (75)$$

$$\dot{C}_2(t) = \beta_2 \sum_{i=1}^{2}\rho_i \int_0^l (x_i-l)\dot{y}_i(x_i,t)\dot{q}_{xi}(x_i,t)dx_i. \quad (76)$$

Then, invoking (23) and (75) in addition to boundary condition, lead to,

$$\dot{C}_1(t) = -\frac{3}{2}\beta_2 \sum_{i=1}^{2} EI_i \int_0^l q_{xxi}^2(x_i,t)dx_i$$
$$- \frac{1}{2}\beta_2 EI_i l q_{xxi}^2(0,t). \quad (77)$$

Substituting $y_i(x_i,t) = q_i(x_i,t) - x_ie_i(t)$ in (76) derives,

$$\dot{C}_2(t) = \beta_2 \sum_{i=1}^{2}\rho_i \int_0^l (x_i-l)\dot{y}_i(x_i,t)\dot{y}_{xi}(x_i,t)dx_i$$
$$+ \beta_2 \sum_{i=1}^{2}\rho_i \int_0^l (x_i-l)\dot{y}_i(x_i,t)\dot{e}_i(t)dx_i. \quad (78)$$

Afterward, by using Lemma 1 along with $\dot{e}(t) = s_i - (\mu_i + c_{3i})z_{3i}$ and boundary condition, we get to,

$$\dot{C}_2(t) \leq -\beta_2 \sum_{i=1}^{2}\rho_i \left\{\frac{1}{2} - l\vartheta_{6i}\right.$$
$$\left. - l(\mu_i + c_{3i})\frac{1}{\vartheta_{7i}}\right\}\int_0^l \dot{y}_i^2(x_i,i)dx_i$$
$$+ \beta_2 l^2 \sum_{i=1}^{2}\rho_i \frac{1}{\vartheta_{6i}}s_i^2 \quad (79)$$
$$+ \beta_2 l^2 \sum_{i=1}^{2}\rho_i(\mu_i + c_{3i})\vartheta_{7i}z_{3i}^2.$$

Substituting (77) and (79) in (74) lead us to,

$$\dot{C}(t) \leq -\frac{3}{2}\beta_2 \sum_{i=1}^{2} EI_i \int_0^l q_{xxi}^2(x_i,t)dx_i$$
$$- \frac{1}{2}\beta_2 EI_i l q_{xxi}^2(0,t)$$
$$- \beta_2 \sum_{i=1}^{2}\rho_i \left\{\frac{1}{2} - l\vartheta_{6i}\right.$$
$$\left. - l(\mu_i + c_{3i})\frac{1}{\vartheta_{7i}}\right\}\int_0^l \dot{y}_i^2(x_i,i)dx_i \quad (80)$$
$$+ \beta_2 l^2 \sum_{i=1}^{2}\rho_i \frac{1}{\vartheta_{6i}}s_i^2$$
$$+ \beta_2 l^2 \sum_{i=1}^{2}\rho_i(\mu_i + c_{3i})\vartheta_{7i}z_{3i}^2.$$

Finally, using (40), (58), (73), and (80), time derivative of $H(t)$ can be derived as,

$$\dot{H}(t) \leq -(\bar{\mu}+c_1)z_1^2 \left\{1 - \frac{\vartheta_1}{(\bar{\mu}+c_1)}\right.$$
$$\left. + \beta_1(EI_1+EI_2)\left\{\frac{1}{2}-\vartheta_3-\frac{1}{\vartheta_2}\right\}\right\}$$
$$- s^2\left\{k - \frac{1}{2}\beta_1(EI_1+EI_2)\right\} - b_2\tilde{m}^2\left(1-\frac{1}{b_3}\right)$$
$$- \gamma_2\tilde{\varepsilon}^2\left(1-\frac{1}{\gamma_3}\right) - \frac{1}{2}a_2\widetilde{W}^T\widetilde{W}$$
$$- \sum_{i=1}^{2}(\mu_i+c_{3i})\left\{1 - \beta_1\frac{1}{\vartheta_{5i}}EI_i - \beta_2 l^2\rho_i\vartheta_{7i}\right\}z_{3i}^2$$
$$- \sum_{i=1}^{2}\left\{k_i - \beta_1 EI_i\vartheta_{4i} - \beta_1 l^2\rho_i\frac{1}{\vartheta_{6i}}\right\}s_i^2 - g_{2i}\tilde{J}_i^2\left(1-\frac{1}{g_{3i}}\right)$$
$$- \zeta_{2i}\tilde{\pi}_i^2\left(1-\frac{1}{\zeta_{3i}}\right) - \frac{1}{2}a_{4i}\widetilde{U}_i^T\widetilde{U}_i$$
$$- \beta_1\sum_{i=1}^{2}EI_i\left\{\frac{1}{2}-\vartheta_2(\bar{\mu}+c_1)\right\}\dot{p}^2(t)$$
$$- \left\{\beta_1(EI_1+EI_2)\left\{\frac{1}{2}-\frac{(\bar{\mu}+c_1)}{\vartheta_3}\right\}\right.$$
$$\left. - \frac{1}{\vartheta_1}\right\}\left(\sum_{i=1}^{2}q_{xxxi}(l,t)\right)^2$$
$$- \sum_{i=1}^{2}EI_i\left\{\frac{1}{2}\beta_2 l - \beta_1\left\{\frac{1}{\vartheta_{4i}}+\vartheta_{5i}(\mu_i+c_{3i})\right\}\right\}q_{xxi}^2(0,t)$$
$$- \frac{3}{2}\beta_2\sum_{i=1}^{2}EI_i\int_0^l q_{xxi}^2(x_i,t)dx_i$$
$$- \beta_2\sum_{i=1}^{2}\rho_i\left\{\frac{1}{2}-l\vartheta_{6i}-l(\mu_i+c_{3i})\frac{1}{\vartheta_{7i}}\right\}\int_0^l \dot{y}_i^2(x_i,i)dx_i$$
$$- \eta|s| - \xi_i|s_i| + \frac{1}{2}a_2 W^{*T}W^* + \gamma_3\gamma_2\bar{\varepsilon}^2 + b_2 b_3 m^2$$
$$+ 0.2785\epsilon_1\bar{\varepsilon} + \frac{1}{2}a_{4i}U_i^{*T}U_i^* + \zeta_{2i}\zeta_{3i}\bar{\pi}_i^2 + g_{2i}g_{3i}\bar{J}_i^2$$
$$+ 0.2785\epsilon_{2i}\bar{\pi}_i \quad \leq -\varphi H(t) + h$$

where, $h = \frac{1}{2}a_2 W^{*T}W^* + \gamma_3\gamma_2\bar{\varepsilon}^2 + b_2 b_3 m^2 + 0.2785\epsilon_1\bar{\varepsilon} + \frac{1}{2}a_{4i}U_i^{*T}U_i^* + \zeta_{2i}\zeta_{3i}\bar{\pi}_i^2 + g_{2i}g_{3i}\bar{J}_i^2 + 0.2785\epsilon_{2i}\bar{\pi}_i$, $\varphi = min\left\{\frac{\kappa_1}{\kappa_2},\frac{\kappa_3}{\kappa_4},\frac{\kappa_5}{\kappa_6}\right\}/\varphi_2$, $\kappa_1 = min\left\{(\bar{\mu}+c_1)\left\{1-\frac{\vartheta_1}{(\bar{\mu}+c_1)}+\beta_1(EI_1+EI_2)\left\{\frac{1}{2}-\vartheta_3-\frac{1}{\vartheta_2}\right\}\right\}, \left\{k-\frac{1}{2}\beta_1(EI_1+EI_2)\right\}, b_2\left(1-\frac{1}{b_3}\right), \gamma_2\left(1-\frac{1}{\gamma_3}\right), \frac{1}{2}a_2\right\}$, $\kappa_2 = max\left\{1,\frac{1}{2a_1},\frac{1}{2\gamma_1},\frac{1}{2b_1},\frac{1}{2}m\right\}$, $\kappa_3 = min\{(\mu_1+c_{31})\left\{1-\beta_1\frac{1}{\vartheta_{51}}EI_1-\beta_2 l^2\rho_1\vartheta_{71}\right\}, (\mu_2+c_{32})\left\{1-\beta_1\frac{1}{\vartheta_{52}}EI_2-\beta_2 l^2\rho_2\vartheta_{72}\right\}, \left\{k_1-\beta_1 EI_1\vartheta_{41}-\beta_1 l^2\rho_1\frac{1}{\vartheta_{61}}\right\}, \left\{k_2-\beta_1 EI_2\vartheta_{42}-\beta_1 l^2\rho_2\frac{1}{\vartheta_{62}}\right\}, g_{21}\left(1-\frac{1}{g_{31}}\right), g_{22}\left(1-\frac{1}{g_{32}}\right), \zeta_{21}\left(1-\frac{1}{\zeta_{31}}\right), \zeta_{22}\left(1-\frac{1}{\zeta_{32}}\right), \frac{1}{2}a_{4i}\}$, $\kappa_4 = max\left\{1,\frac{1}{2a_{31}},\frac{1}{2\zeta_{11}},\frac{1}{2g_{11}},\frac{1}{2}J_1,\frac{1}{2a_{32}},\frac{1}{2\zeta_{12}},\frac{1}{2g_{12}},\frac{1}{2}J_2\right\}$, $\kappa_5 = min\{\beta_2\rho_1\left\{\frac{1}{2}-l\vartheta_{61}-l(\mu_1+c_{31})\frac{1}{\vartheta_{71}}\right\}, \beta_2\rho_2\left\{\frac{1}{2}-l\vartheta_{62}-l(\mu_2+c_{32})\frac{1}{\vartheta_{72}}\right\}, \frac{3}{2}\beta_2 EI_1, \frac{3}{2}\beta_2 EI_2\}$, $\kappa_6 = \frac{\beta_1}{2}max\{EI_1,EI_2,\rho_1,\rho_1\}$.

Moreover, following conditions hold,

$$1 - \frac{\vartheta_1}{(\bar{\mu}+c_1)} + \beta_1(EI_1+EI_2)\left\{\frac{1}{2}-\vartheta_3-\frac{1}{\vartheta_2}\right\} > 0 \quad (82)$$

$$k - \frac{1}{2}\beta_1(EI_1+EI_2) > 0,$$
$$k_i - \beta_1 EI_i\vartheta_{4i} - \beta_1 l^2\rho_i\frac{1}{\vartheta_{6i}} > 0 \quad (83)$$

$$1 - \frac{1}{b_3} > 0, 1 - \frac{1}{\gamma_3} > 0, 1 - \frac{1}{g_{3i}} > 0, 1 - \frac{1}{\zeta_{3i}} > 0 \quad (84)$$

$$1 - \beta_1\frac{1}{\vartheta_{5i}}EI_i - \beta_2 l^2\rho_i\vartheta_{7i} > 0 \quad (85)$$

$$\beta_1(EI_1+EI_2)\left\{\frac{1}{2}-\frac{(\bar{\mu}+c_1)}{\vartheta_3}\right\} - \frac{1}{\vartheta_1} > 0 \quad (86)$$

(81)
$$\frac{1}{2}\beta_2 l - \beta_1\left\{\frac{1}{\vartheta_{4i}}+\vartheta_{5i}(\mu_i+c_{3i})\right\} > 0 \quad (87)$$

$$\frac{1}{2} - l\vartheta_{6i} - l(\mu_i+c_{3i})\frac{1}{\vartheta_{7i}} > 0,$$
$$\frac{1}{2} - \vartheta_2(\bar{\mu}+c_1) > 0. \quad (88)$$

Now, from (81), the following holds,

$$\frac{d(H(t)e^{\varphi t})}{dt} \leq he^{\varphi t} \quad (89)$$

which results in,

$$H(t) \leq H(0)e^{-\varphi t} + \frac{h}{\varphi}. \quad (90)$$

Therefore, invoking (41), (44), (47), (71), and (90), lead to,

$$\frac{1}{2}z_1^2(t) \leq \frac{2H(t)}{\varphi_2\varphi_4} \leq \frac{2}{\varphi_2\varphi_4}\left\{H(0)e^{-\varphi t}+\frac{h}{\varphi}\right\} \quad (91)$$

that means when $t \to \infty$,

$$|z_1(t)| \leq \sqrt{\frac{2h}{\varphi_2\varphi_4\varphi}}, \quad (92)$$

where $\varphi_4 = min\left\{1,\frac{1}{2a_1},\frac{1}{2\gamma_1},\frac{1}{2b_1},\frac{1}{2}m\right\}$. In the same way, combining (59), (62), (65), (71), and (90), we get to,

$$\frac{1}{2}z_{3i}^2(t) \leq \frac{2H(t)}{\varphi_2\varphi_{5i}} \leq \frac{2}{\varphi_2\varphi_{5i}}\left\{H(0)e^{-\varphi t}+\frac{h}{\varphi}\right\} \quad (93)$$

where $i = 1,2$, and, when $t \to \infty$,

$$|z_{3i}(t)| \leq \sqrt{\frac{2h}{\varphi_2\varphi_{5i}\varphi}}, \quad (94)$$

where, $\varphi_{5i} = min\left\{1,\frac{1}{2a_{3i}},\frac{1}{2\zeta_{1i}},\frac{1}{2g_{1i}},\frac{1}{2}J_i\right\}, i = 1,2$. Based on (92) and (94), it can be concluded that $y_M(t)$ and $\theta_i(t), i = 1,2$, ultimately converge to $y_M^d$ and $\theta_i^d$, respectively. Therefore, by accomplishing the motion control objectives, the desired contact force will be applied to the object smoothly and ultimately.

## IV. NUMERICAL SIMULATIONS

In this paper, in order to accomplishing force control task, NABFC law is proposed to overcome unknown model uncertainties and unknown input constraints. The hybrid PDE-ODE equations of the system (8)-(11), driven by use of Hamilton principle, alongside constraint equation (12) are used to derive control and update laws (36)-(39) and (54)-(57). In this section, to show the performance of the designed controller, numerical solution is employed. The parameters of the dual one-link flexible arms are given by: $EI_i = 0.115\ N.m^2, \rho_i = 0.054\frac{kg}{m}$, $M = 0.3\ kg, m_i = 0.1\ kg, l = 0.2\ m, J_i 0.0073\frac{Kg}{m^2}, L = 0.1\ m$, where $i = 1,2$. The saturation levels $k_{M1} = 2.1, k_{m1} = -2.1, k_{M2} = 0.31, k_{m2} = -0.31$, the dead-zones ranges $m_{r1} = 0.1, m_{l2} = -0.1, m_{r2} = 0.01, m_{l2} = -0.01$, and the slope parameters $k_{li} = 1, k_{ri} = 1, i = 1,2$. Moreover, the model uncertainties are supposed as $\Delta_i = -0.1 EI_i \dot{w}_{xxi}(0,t), \bar{\Delta} = -0.05(EI_1 + EI_2)(\dot{w}_{xxx1}(l,t) + \dot{w}_{xxx2}(l,t))$. Further, the reference signals for the contact force is $\lambda_1^d = -0.5N$ and for angular displacement is $\theta_1^d = 3\ deg$, where other values can be derived from (16). In order to avoid chattering phenomena, the sign function in control laws is replaced by sat function, which can be defined as below,

$$sat(\iota) = \begin{cases} \iota_{max} & if\ \iota \geq \iota_{max} \\ \iota & if\ \iota_{max} < \iota < \iota_{min} \\ \iota_{min} & if\ \iota \leq \iota_{min} \end{cases}$$

where $\iota_{max} = -\iota_{min} = 0.01$.

The initial values of this constrained system must be chosen in a way that constraint equation (12) be satisfied. For this reason, we set initial values as, $\theta_i(0) = \dot{\theta}_i(0) = 0, i = 1,2, and\ y_M(0) = 0.05, \dot{y}_M(0) = 0$.

*Scenario I*: With the proposed NABFC.

The design parameters of the NABFC scenario are chosen in a way that satisfy conditions (82)-(88).

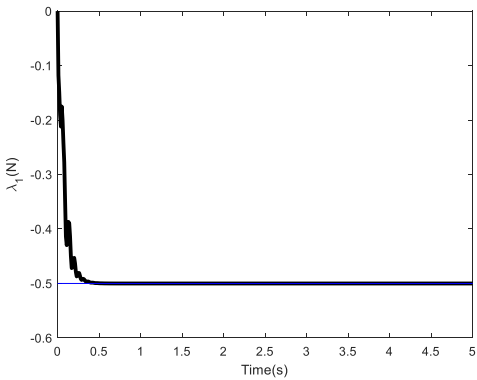

Fig. 2. Force response of NABFC for first link

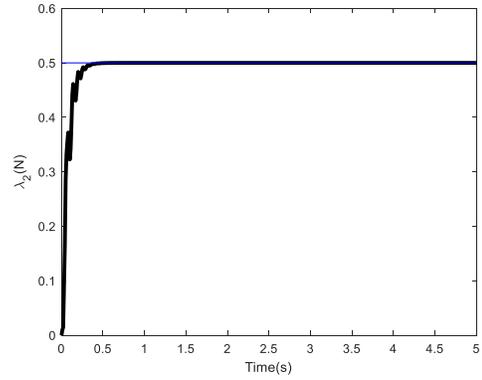

Fig. 3. Force response of NABFC for second link

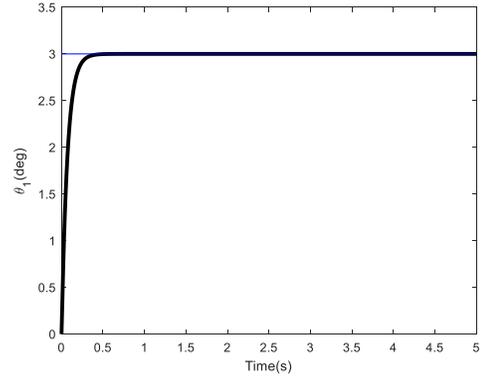

Fig. 4. NABFC response fore angular displacement of first link

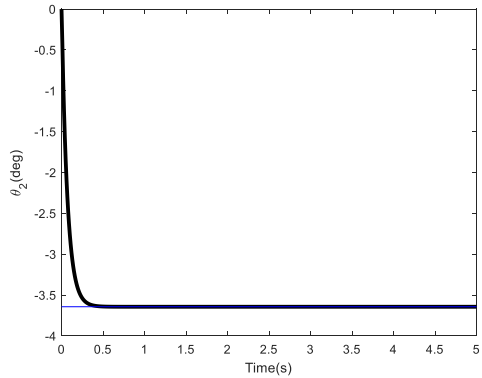

Fig. 5. NABFC response fore angular displacement of second link

As shown from Fig. 2 and Fig. 3, under the proposed NABFC laws, the contact force between end-effectors and grasped object ultimately converged to the desired value. The excellent performance of the controller can be inferred from convergence time of less than 0.5 second. Having slight fluctuation before reaching the set-point, which is inflicted by input constraints, the desired force is applied successfully without any overshoot, which prepares a safe grasping task for fragile objects. Further thought, the proposed controller canceled all model and input uncertainties and performed flawlessly.

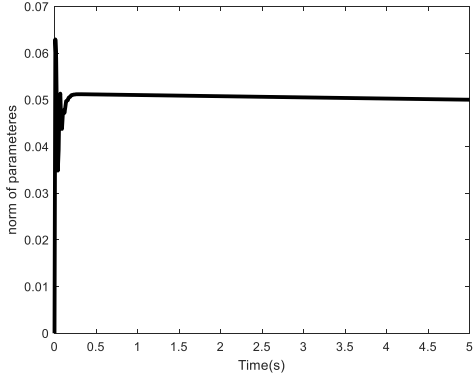

Fig. 6. Vector norm of estimated parameters

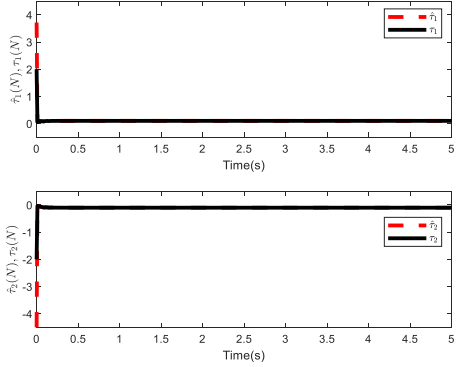

Fig. 7. NABFC control signals at the root

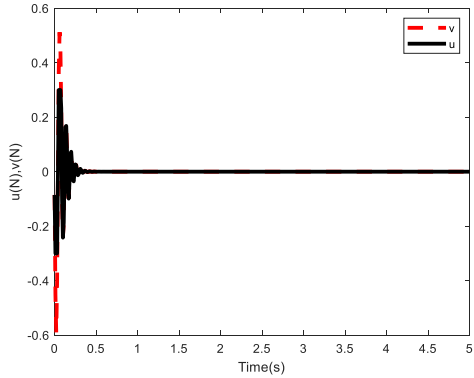

Fig. 8. NABFC control signal at the end-effector

Angular displacement of each link is shown in Fig. 4 and Fig. 5. As it is concludable, controller was able to perfectly relocate links to the desired position at less than 0.5 second. This consistency in the convergence time perfectly proves two things: correctness of the assumption that by achieving position control problem goals force control problem objectives will be established, and stability analysis result. Fig. 6 demonstrates that the norm of parameter estimation converged to a fix point. One of the abilities of the adaptive control is to accomplish control objective despite inaccurate estimation of unknown parameters. Therefore, by achieving the position control objectives, the desired force applied to the object with no need for force sensor or force control approaches. Fig. 7 and Fig. 8, however, depict control signals of two motor torques at the root and one force at the end-effector, which solid line represents actuator output and dotted line denotes designed controller. As it can be seen, in spite of input constraint and less motor power, controller could afford to exhibit such a great performance. Therefore, results illustrate that controller clearly overcome the uncertainty and input saturation and ultimately accomplished the position control goals, which implicates correctness of the stability analysis outcomes.

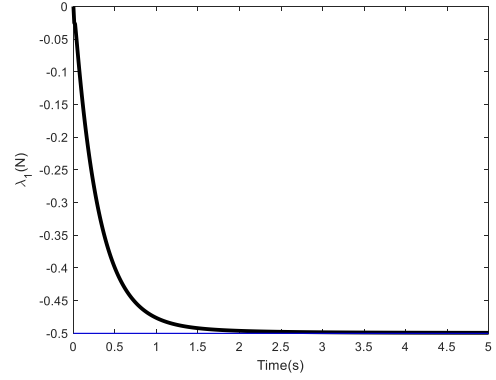

Fig. 9. Force response of PDS control for first link

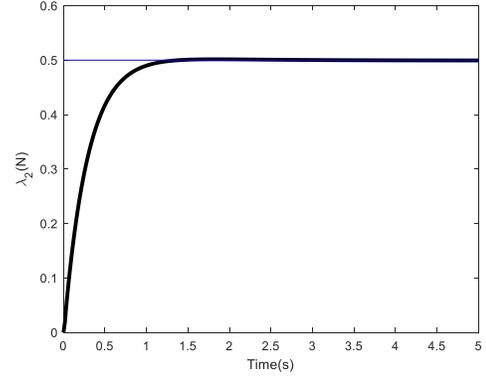

Fig. 10. Force response of PDS control for second link

*Scenario II*: PDS control.

The PDS control signal is defined as below:

$$v_{PDS}(t) = -\bar{k}_p p(t) - \bar{k}_v \dot{p}(t) + (EI_1 w_{xxx1}(l,t) + EI_2 w_{xxx2}(l,t)) \quad (95)$$

$$\tau_{PDS_i} = -\bar{k}_{p_i} e_i(t) - \bar{k}_{v_i} \dot{e}_i(t) - \bar{k}_{s_i} EI_i q_{xxi} + EI_i w_{xxi}(0,t) \quad (96)$$

where $\bar{k}_p, \bar{k}_v, \bar{k}_{p_i}, \bar{k}_{v_i}, \bar{k}_{s_i}, i = 1,2$, are positive constants and as the design parameters of the PD controller, their values are: $\bar{k}_p = 40, \bar{k}_v = 35, \bar{k}_{p_i} = 35, \bar{k}_{v_i} = 15, \bar{k}_{s_i} = 10$.

Both Fig. 9 and Fig. 10 display time history of contact forces. As it can be seen, it takes almost 2 second for PDS controller to apply desired force to the object. Fig. 11 and Fig. 12 illustrate that the same time was required for the controller to accomplish position control objectives. Based on Fig. 13 and Fig. 14, which exhibit the control signals of PDS, it can be concluded that for the same input power, NABFC is much faster yet robust and stable than PDS controller. This clearly shows the excellent performance of the proposed controller.

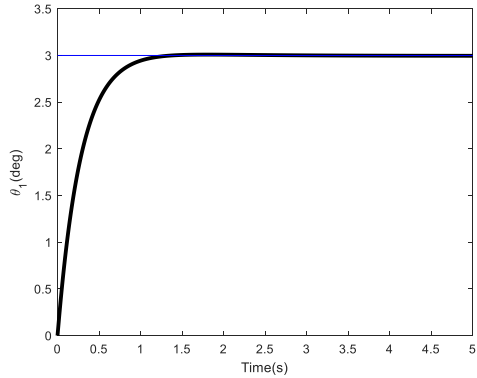

Fig. 11. PDS response for angular displacement of first link

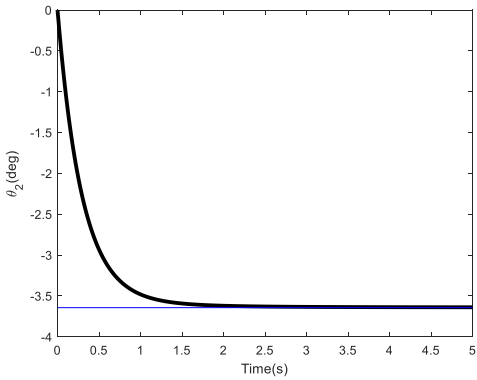

Fig. 12. PDS response for angular displacement of second link

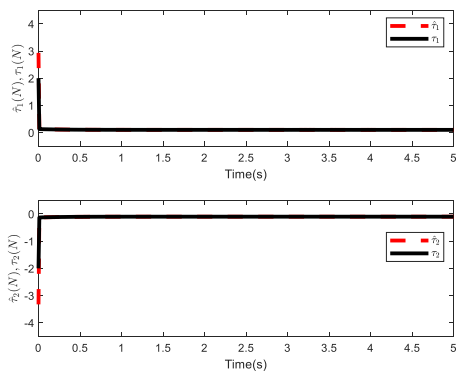

Fig. 13. PDS Control signals at the root

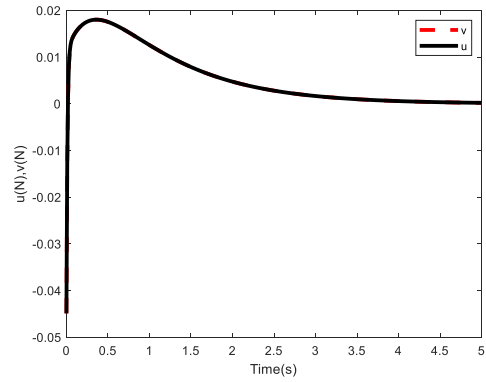

Fig. 14. PDS control signal at the end-effector

*Scenario III*: PD control.
The PD controller is introduced as following:

$$v_{PD}(t) = -k_p p(t) - k_v \dot{p}(t) + (EI_1 w_{xxx1}(l,t) + EI_2 w_{xxx2}(l,t)) \tag{97}$$

$$\tau_{PD_i} = -k p_i e_i(t) - k_{v_i} \dot{e}_i(t) + EI_i w_{xxi}(0,t) \tag{98}$$

where $k_p, k_v, k_{p_i}, k_{v_i}, i = 1,2$, are positive constants and as the design parameters of the PD controller, their values are: $k_p = 10, k_v = 9, k_{p_i} = 60, k_{v_i} = 55$.

Fig. 15 and Fig. 16 show the time history of contact force applied to the object by PD controller. As it can be observed, it took much longer for PD controller to achieve goals, although the input power is the same as NABFC and PDS control. The reason that PDS is much better than PD is due to strain feedback in the PDS that makes it faster. However, all results perfectly evidence that the introduced NABFC controller is robust in canceling uncertainties and fast in convergence.

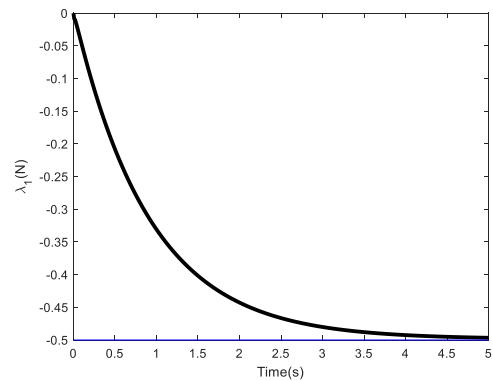

Fig. 15. Force response of PD control for first link

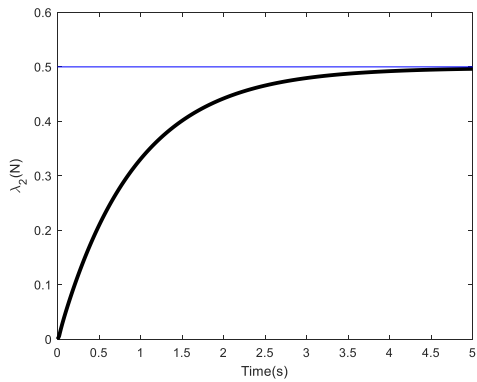

Fig. 16. Force response of PD control for second link

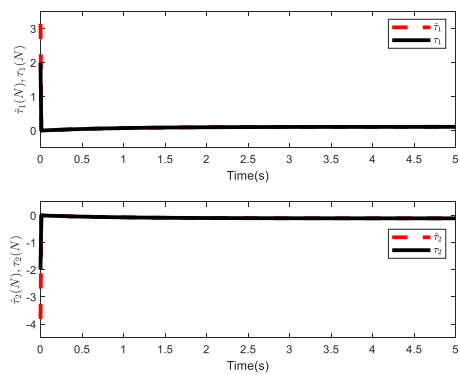

Fig. 17. PD Control signals at the root

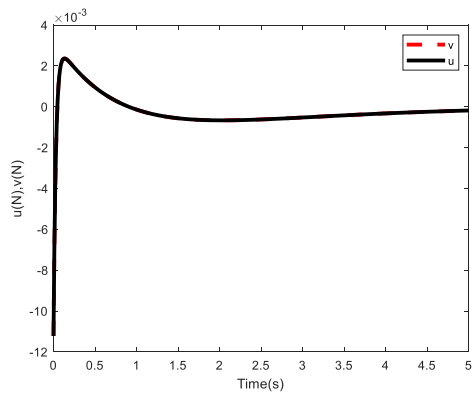

Fig. 18. PD control signal at the end-effector

## Conclusion

In this article, we proposed the NABFC to control contact force between end-effectors and grasped object. For this aim and in order to remove force sensor, we utilized the constraint of the system and turned the force control problem to position control problem. Afterwards, we proposed a systematic, Lyapunov-based stability analysis procedure that removes the prevailing problems of approaches utilized in the previous works. By doing so, we were able to consider unknown, mixed input saturation and input dead zones, along with unknown model and parameters uncertainties for the first time in the field of dual one-link flexible arms. Then, to handle these challenges, a combination of intelligent control schemes, Neural Networks, with robust, systematic approaches like backstepping-sliding mode, was exploited. Through the stability analysis by using the proposed Lyapunov-based approach, uniform ultimate boundedness was achieved. Finally, numerical simulation was used to demonstrate excellent performance of the controller. As figures have shown, proposed controller can afford to cancel all uncertainties, and, despite input constraints, uniformly apply desired force to object in 0.5 second. By comparing result of the NABFC with PDS and PD control, it was proved that the proposed controller is faster in convergence and robust in canceling uncertainties, for the same input power.

References

[1] H.sayyaadi and M.Hejrati, "Adaptive Backsteping Sliding Mode control of single Link Flexible Manipulator in Presence of State constraint and nonlinear observer," *4th Int. Conf. Robot. Mechantronics(ICRoM)*, 2019.

[2] Y. Ren, Z. Zhao, C. Zhang, Q. Yang, and K.-S. Hong, "Adaptive Neural-Network Boundary Control for a Flexible Manipulator With Input Constraints and Model Uncertainties," *IEEE Trans. Cybern.*, pp. 1–12, 2020, doi: 10.1109/tcyb.2020.3021069.

[3] K. Yamaguchi, T. Endo, Y. Kawai, and F. Matsuno, "Non-collocated boundary control for contact-force control of a one-link flexible arm," *J. Franklin Inst.*, vol. 357, no. 7, pp. 4109–4131, 2020, doi: 10.1016/j.jfranklin.2020.01.018.

[4] W. He, T. Meng, X. He, and C. Sun, "Iterative learning control for a flapping wing micro aerial vehicle under distributed disturbances," *IEEE Trans. Cybern.*, vol. 49, no. 4, pp. 1524–1535, 2019, doi: 10.1109/TCYB.2018.2808321.

[5] H. Sayyaadi and M. Hejrati, "Boundary Force Control of Two one-link Flexible Manipulator to Accomplish Safe Grasping Task," 2021, [Online]. Available: https://civilica.com/doc/1238334/.

[6] H. Talebi, R. Patel, and K. Khorasani, *Control of flexible-link manipulators using neural networks*. .

[7] J. Liang *et al.*, "Dual quaternion based kinematic control for Yumi dual arm robot," in *2017 14th International Conference on Ubiquitous Robots and Ambient Intelligence, URAI 2017*, Jul. 2017, pp. 114–118, doi: 10.1109/URAI.2017.7992899.

[8] A. Zhai, J. Wang, H. Zhang, G. Lu, and H. Li, "Adaptive robust synchronized control for cooperative robotic manipulators with uncertain base coordinate system," *ISA Trans.*, Jul. 2021, doi: 10.1016/j.isatra.2021.07.036.

[9] Z. Li, Y. Xia, and F. Sun, "Adaptive fuzzy control for multilateral cooperative teleoperation of multiple robotic manipulators under random network-induced delays," *IEEE Trans. Fuzzy Syst.*, vol. 22, no. 2, pp. 437–450, 2014, doi: 10.1109/TFUZZ.2013.2260550.

[10] T. Endo, F. Matsuno, and H. Kawasaki, "Simple boundary cooperative control of two one-link flexible arms for grasping," *IEEE Trans. Automat. Contr.*, vol. 54, no. 10, pp. 2470–2476, 2009, doi: 10.1109/TAC.2009.2029401.

[11] T. Endo, F. Matsuno, and Y. Jia, "Boundary cooperative control by flexible Timoshenko arms," *Automatica*, vol. 81, pp. 377–389, 2017, doi: 10.1016/j.automatica.2017.04.017.

[12] T. Endo, D. Wu, and F. Matsuno, "Boundary control of dual Timoshenko arms for grasping and orientation control," *Int. J. Control*, vol. 0, no. 0, pp. 1–11, 2018, doi: 10.1080/00207179.2018.1514532.

[13] T. Endo, N. Shiratani, K. Yamaguchi, and F. Matsuno, "Grasp and Orientation Control of an Object by Two Euler–Bernoulli Arms With Rolling Constraints," *J. Dyn. Syst. Meas. Control*, vol. 141, no. 12, 2019, doi: 10.1115/1.4044718.

[14] T. Endo, K. Umemoto, and F. Matsuno, "Exponential stability of dual

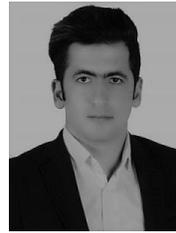

**Mahdi Hejrati** was born in Tabriz, East Azerbaijan, Iran, in 1996. He received the B.S. degree in mechanical engineering from the University of Bonab, East Azerbaijan, in 2018, and the M.S. degree in mechanical engineering from Sharif University of Technology (SUT), Tehran, in 2021.

During the M.S. program, from 2018 to 2019, he cooperated with the Center of Excellence in Design, Robotics, and Automation (CEDRA), Sharif University of Technology. Moreover, from 2019 to 2020, he was a member of the engineering team to design and manufacture a soft knee exosuit for gait rehabilitation in stroke in cooperation with the New Technologies Research Center, Amir Kabir University of Tehran. His research interests include flexible manipulators, robotics, control theory, and mechatronics.